\documentclass[12pt]{article}
\usepackage{graphicx,amssymb,amsfonts,amsmath,epsfig}

\makeatletter
\DeclareSymbolFont{AMSb}{U}{msb}{m}{n}
\DeclareSymbolFontAlphabet{\mathbb}{AMSb}
\renewcommand{\section}{\@startsection{section}{1}{\z@}%
                                    {-7ex \@plus -1ex \@minus -.2ex}%
                                    {2.5ex \@plus.2ex}%
                                    {\normalfont\large\scshape\centering}}
\renewcommand{\subsection}{\@startsection{subsection}{2}{\z@}%
                                       {-5ex \@plus -1ex \@minus -.2ex}%
                                       {1.5ex \@plus.2ex}%
                                       {\normalfont\normalsize\scshape}}
\renewcommand{\subsubsection}{\@startsection{subsubsection}{3}{\z@}%
                                       {-5ex \@plus -1ex \@minus -.2ex}%
                                       {1.5ex \@plus.2ex}%
                                       {\normalfont\normalsize\scshape}}

\renewcommand\@seccntformat[1]{\ignorespaces\csname #1name\endcsname\space
                               \csname the#1\endcsname.\quad}   
\newdimen\captionmargin
\newdimen\captionindent
\newdimen\captionwidth
\newcommand{\captionfont}{\slshape}
\newcommand\@captionlabel[1]{\textsc{#1:}\space}
\long\def\@makecaption#1#2{%
  \vskip\abovecaptionskip
  \captionwidth\hsize
  \advance\captionwidth -2\captionmargin
  \sbox\@tempboxa{\@captionlabel{#1}\captionfont #2}%
  \ifdim \wd\@tempboxa >\captionwidth
    \ifdim\captionindent>\z@
      \advance\captionwidth -\captionindent
      \hskip\captionindent
    \fi
    \hskip\captionmargin
    \parbox[t]{\captionwidth}{\leavevmode\hskip-\captionindent
      \@captionlabel{#1}\captionfont #2}%
  \else
    \global \@minipagefalse
    \hb@xt@\hsize{\hfil\box\@tempboxa\hfil}%
  \fi
  \vskip\belowcaptionskip}
\def\eqnarray{%
   \stepcounter{equation}%
   \def\@currentlabel{\p@equation\theequation}%
   \global\@eqnswtrue
   \m@th
   \global\@eqcnt\z@
   \tabskip\@centering
   \let\\\@eqncr
   $$\everycr{}\halign to\displaywidth\bgroup
       \hskip\@centering$\displaystyle\tabskip\z@skip{##}$\@eqnsel
      &\global\@eqcnt\@ne$\;\hfil{##}$\hfil
      &\global\@eqcnt\tw@$\;\displaystyle{##}$\hfil\tabskip\@centering
      &\global\@eqcnt\thr@@ \hb@xt@\z@\bgroup\hss##\egroup
         \tabskip\z@skip
      \cr}
\ifcase \@ptsize
\or
\or
\fi
  \divide\@tempdima\baselineskip
  \@tempcnta=\@tempdima
\makeatother
\begin{document}

%
%

\renewcommand{\theequation}{\arabic{section}.\arabic{equation}}
\renewcommand{\thefigure}{\arabic{figure}}
\newcommand{\gapprox}{%
\mathrel{%
\setbox0=\hbox{$>$}\raise0.6ex\copy0\kern-\wd0\lower0.65ex\hbox{$\sim$}}}
\textwidth 165mm \textheight 220mm \topmargin 0pt \oddsidemargin 2mm
\def\ib{{\bar \imath}}
\def\jb{{\bar \jmath}}

\newcommand{\ft}[2]{{\textstyle\frac{#1}{#2}}}
\newcommand{\be}{\begin{equation}}
\newcommand{\ee}{\end{equation}}
\newcommand{\bea}{\begin{eqnarray}}
\newcommand{\eea}{\end{eqnarray}}
\newcommand{\Identity}{{1\!\rm l}}
\newcommand{\cx}{\overset{\circ}{x}_2}
\def\CN{$\mathcal{N}$}
\def\CH{$\mathcal{H}$}
\def\hg{\hat{g}}
\newcommand{\bref}[1]{(\ref{#1})}
\def\espai{\;\;\;\;\;\;}
\def\zespai{\;\;\;\;}
\def\avall{\vspace{0.5cm}}
\newtheorem{theorem}{Theorem}
\newtheorem{acknowledgement}{Acknowledgment}
\newtheorem{algorithm}{Algorithm}
\newtheorem{axiom}{Axiom}
\newtheorem{case}{Case}
\newtheorem{claim}{Claim}
\newtheorem{conclusion}{Conclusion}
\newtheorem{condition}{Condition}
\newtheorem{conjecture}{Conjecture}
\newtheorem{corollary}{Corollary}
\newtheorem{criterion}{Criterion}
\newtheorem{defi}{Definition}
\newtheorem{example}{Example}
\newtheorem{exercise}{Exercise}
\newtheorem{lemma}{Lemma}
\newtheorem{notation}{Notation}
\newtheorem{problem}{Problem}
\newtheorem{prop}{Proposition}
\newtheorem{rem}{{\it Remark}}
\newtheorem{solution}{Solution}
\newtheorem{summary}{Summary}
\numberwithin{equation}{section}
\newenvironment{pf}[1][Proof]{\noindent{\it {#1.}} }{\ \rule{0.5em}{0.5em}}
\newenvironment{ex}[1][Example]{\noindent{\it {#1.}}}

\thispagestyle{empty}

\begin{flushright}\scshape
January 2004
\end{flushright}
\vskip1cm

\begin{center}

{\LARGE\scshape Truncations driven by constraints:
consistency and conditions for correct upliftings
\par}
\vskip15mm

\textsc{Josep M. Pons$^{a}$ and Pere Talavera$^{b}$}
\par\bigskip
$^a${\em
Departament d'Estructura i Constituents de la Mat\`eria,
Universitat de Barcelona,\\
Diagonal 647, E-08028 Barcelona, Spain.}\\[.1cm]
$^b${\em
Departament de F{\'\i}sica i Enginyeria Nuclear,
Universitat Polit\`ecnica de Catalunya,\\
Jordi Girona 1--3, E-08034 Barcelona, Spain.}\\[.1cm]
\vspace{5mm}
\end{center}

\section*{Abstract}
We discuss the mechanism of truncations driven by the
imposition of constraints. We show how the consistency of such truncations is
controlled, and give general theorems that establish conditions for the correct
uplifting of solutions. We show in some particular examples how one can get correct
upliftings from $7d$ supergravities to $10d$ type IIB supergravity, even in cases when
the truncation is not initially consistent by its own.

\vspace{3mm} \vfill{ \hrule width 5.cm \vskip 2.mm {\small
\noindent E-mail: pons@ecm.ub.es, pere.talavera@upc.es }}

\newpage
\setcounter{page}{1}


\tableofcontents       %
\vskip 1cm             %

\setcounter{equation}{0}

\section{Motivation and general set up}
\label{sec:intro}

In recent years, owing to the success of the AdS/CFT
correspondence, there has been intense work in trying to extend
this correspondence to supergravity backgrounds preserving less
than maximal supersymmetry. One way of achieving the construction
of such models is to derive them as upliftings of solution of
lower dimensional theories, related to the higher dimensional ones
by a process of \emph{truncation}. The physical motivation in terms 
of the ground state can be seen as follows: gravitational theories in higher
dimension can lead after spontaneous compactification to theories
that accommodate the Standard Model or the like at $4d$
\cite{Manton:1979kb}\footnote{A possible way to determine the ground state 
is by an uplifting procedure.}. 
Thus the higher dimensional theory can be
seen as a ``unification'' theory. Even though the approach is
appealing it has non-trivial difficulties and after some decades
of work is still lacking a general procedure.

The scope of this paper is not to 
present a new procedure
of \emph{truncation} nor a new solution for supergravity
backgrounds preserving some amount of supersymmetry. Our only
concern here is just to systematize a general procedure for
obtaining upliftable solutions for certain cases of truncations.
Several results in this respect are presented under the form of
theorems. In addition, we give some examples of models currently 
used in the literature to enlighten their construction and the consistency 
of the truncations involved. 

Before proceeding let us define briefly the concept of
\emph{truncation}, which has been already used above. 
This definition will be expanded later on. 
This word, \emph{truncation}, is
commonly used with two different, though related, meanings. 
Given an action
functional on a manifold describing a theory one says, {\it grosso
modo}, that it has undergone a truncation if: {\it i)} there
has been a reduction of the space-time coordinates --that is, a
dimensional reduction--, or {\it ii)} the number of d.o.f. in the
theory (fields or field components) has been reduced.

\avall

In
the remainder we shall describe briefly these two types of truncation.

\avall

\begin{itemize}
\item[{\it{i)}}] First-type truncation, or pure dimensional
reduction. In the most simplest setting the dimension
of the space-time is reduced by considering the description,
through a new action principle, of the
subset of solutions of the original equations of motion (e.o.m.)
that share some Killing symmetries, while keeping unchanged the
number of d.o.f.  attached to every space-time point.
In more general terms, dimensional reductions proceed along these two main
categories
\begin{itemize}
\item {\it The reduction is performed via a group manifold of
independent Killing symmetries}. 
Its actual form is dictated by the isometry group of
the field configurations. The reduction is
consistent\footnote{The concept of consistency for a truncation
will be defined below.} as long as the
tracelessness condition is fulfilled by the 
structure constants of the Lie group 
\cite{Scherk:1979zr,Maccallum:gd,pere-pep}.

\item {\it The reduction is done on a coset space}. In
this case the Killing vectors are no longer independent.
Despite many efforts we believe that the state of the art nowadays
does not provide yet with a systematic understanding of
this kind of truncation procedure and its consistency.
{F}or a review up to the 80's see for instance \cite{Duff:hr} (and references
therein).
Recent developments can be found in 
\cite{Cvetic:2003jy,Nastase:1999kf}.

In this case the reduction is performed under some symmetry
considerations but usually for its own consistency
needs to be mixed with a second-type truncation (see below) on the 
field configurations. 

\end{itemize}
\item[{\it{ii})}] Second-type truncation. It
consists in the introduction of constraints that produce a further
reduction of the number of independent fields --or field
components-- defining the theory. We shall only consider
constraints in configuration space. Let us mention that, as
will be shown below, the consistency of this type of
truncations is model dependent and ought to be considered in a
case a case basis.
\end{itemize}

As has been mentioned before these two types of
truncation are usually applied altogether under the common concept
of dimensional reduction, but in order to give some insight
on the model construction we think it is very convenient to
maintain a clear distinction between them. Notice, however,
that in both cases we are producing a {\sl truncation} in the
field content of the theory, either because an infinity of
Kaluza-Klein modes are eliminated in the dimensional reduction
process or because some field components become redundant due to
the presence of constraints.

\avall

The results presented above concerning the consistency of a
truncation are purely classical. One can of course use these results
within a strategy to obtain solutions of the higher dimensional
theory by uplifting from solutions, perhaps easier to find, of the lower
dimensional one. These solutions can be for instance candidates 
for a ground state.

A different and in some sense complementary perspective is that of
compactification \cite{Witten:me,Duff:1983vj}. It already considers a classical 
solution, for instance, a vacuum undergoing spontaneous compactification (as a
spontaneous symmetry breaking) that exhibits a space-time with a 
compact component, and formulates a quantum field theory on
this background. An expansion in modes (Kaluza-Klein) over the compact
space may allow to keep only the massless modes in an effective field
theory sense, thus ending up with an effective truncation of the theory.

\avall

Before proceeding we shall state the concept of \emph{consistent
truncation}. If we
denote generically as $\Phi$ the original fields, and $\tilde\Phi$
the remaining fields after the truncation, a definite prescription
allows the reconstruction of the configuration of the former from
that of the latter. At the level of the variational principle
there is a natural map from an initial Lagrangian ${\mathcal
L}^{(d+n)}$ into a new one ${\mathcal L}^{(d)}$, perhaps with
lower dimensionality --when $n
>0$--, perhaps with fewer d.o.f.\,,
\begin{equation}
\label{runcation} 
{\mathcal L}^{(d+n)}\left(\Phi({\bf x},{\bf y})\right) \rightarrow {\mathcal
L}^{(d)}\left(\tilde\Phi({\bf x})\right)\,,
\end{equation}
with $\Phi$ ($\tilde\Phi$) any field of the original
(final) theory. Then {\it a truncation, be it first-type,
second-type or mixed, is consistent if the solutions of the
equations of motion for the reduced Lagrangian ${\mathcal L}^{(d)}$
are still solutions of the e.o.m. for the original Lagrangian
${\mathcal L}^{(d+n)}$}.

The issue of consistency for second-type
truncations was examined in \cite{pere-pep} but the answer given there was
incomplete, for, although some particular cases were worked out,
no general result was given concerning the effect of the
introduction of constraints. It is our purpose to amend here this
incompleteness.

\section{Second-type truncations. Dynamical consequences
of constraints} \label{general}

In this section we shall give general results on second-type
truncations. The introduction of constraints (denoted
by $f^A(\Phi)$, where $\Phi$ represents generically the
fields in the theory) on the configuration space of a field theory
may have two distinct effects, according to how they affect the
gauge freedom of the theory. They can act as gauge-fixing
constraints, thus restricting the gauge freedom, or they may
respect the gauge freedom. The first case has been dealt with in
\cite{plugging} and will not be discussed further. 
The second case will be our concern here.

\avall

The reduced theory is obtained after inserting the
conditions $f^A=0$ into the original Lagrangian. The typical
way to proceed is to find an independent set of fields, 
denoted generically as $\tilde{\Phi}$, out of the original set
$\Phi$, such that the constraints $f^A=0$ can be equivalently
written as $\Phi = F(\tilde{\Phi})$. We represent it as
$$
{\mathcal L} \longrightarrow {\mathcal L}_{\rm R}:=({\mathcal L})_{\!f^A=0}
\quad {\rm with}~~ {\mathcal L}_{\rm R}[\tilde{\Phi}] := {\mathcal L}[F(\tilde{\Phi})]\,.
$$

Using the Dirac-Bergman approach to deal with the formalism of gauge
theories \cite{dirac2,bergm3}, we observe that in the canonical formalism there will
exist some primary constraints $\phi_\mu$ that, unlike the
constraints $f^A=0$ introduced ad hoc, are inherently born to the
formalism. In order to respect the gauge freedom, we shall assume that {\it the
truncation constraints are first-class with respect to the primary
constraints}. Technically this corresponds to the Poisson bracket
\begin{equation}
\{\phi_\mu\,,f^A \} \approx 0\,,\label{assumpt}
\end{equation}
i.e., $\{\phi_\mu\,,f^A \}$ vanishes on the surface defined by
$\phi_\mu=0$\,, $f^A=0$.

With this assumption the following theoretical result
can be formulated 

\vspace{6mm}

\noindent {\bf Theorem 1:} {\sl The e.o.m. of ${\mathcal L}$,
in addition to the condition to satisfy the constraints $f^A=0$,
are equivalent to the e.o.m. of  ${\mathcal L}_{\rm R}$ plus the
condition to satisfy some --secondary-- constraints $\chi^A=0$}\,.

\vspace{6mm}

We shall only sketch the proof of the theorem, leaving the complete,  
technical proof, together with
the precise construction of the constraints $\chi^A$, to the appendix. 
The condition for the dynamics defined by the Lagrangian 
${\mathcal L}$ to be compatible with the constraints $f^A=0$, 
which is a tangency condition, imposes in general 
the existence of new, secondary, constraints. When the constraints $f^A=0$ are
plugged within these secondary constraints we get the constraints $\chi^A=0$.
Then one can prove that the dynamics defined by ${\mathcal L}$ on $f^A=0$ 
coincides with the dynamics defined by ${\mathcal L}_{\rm R}$ on $\chi^A=0$.

Notice that this theorem guarantees that any solution of the
reduced theory ${\mathcal L}_{\rm R}$ satisfying the secondary constraints
$\chi^A=0$ can be uplifted to a solution of the original theory
${\mathcal L}$ satisfying $f^A=0$. On the other hand, if a solution of
${\mathcal L}_{\rm R}$ does not satisfy the constraints $\chi^A=0$ it will
not be upliftable to a solution of the original theory ${\mathcal L}$.

The presence of these new constraints $\chi^A=0$ in the reduced
theory is bound to make, in principle, the truncation
inconsistent, except for some exceptional cases where the
restrictions $\chi^A=0$ are void or already included in the e.o.m.
of the reduced Lagrangian. But we have just opened the way for
a {\sl stabilization mechanism}: we can start again with the new
Lagrangian ${\mathcal L}_{\rm R}$ and the new constraints $\chi^A=0$ and
perform another truncation of the same type. Several possibilities are open
when we try to run this stabilization mechanism again. 
\begin{itemize}
\item[{\it{i)}}]
It
might be that some of the constraints $\chi^A$ are indeed gauge
fixing constraints. As we have said before, their effect on the
theory, see \cite{plugging}, is quite different from the one just
examined and must be dealt with accordingly. 
\item[{\it{ii)}}]
Some of the
constraints $\chi^A$ may not be holonomic, so the process of
reducing the d.o.f.  may become more sophisticated than
just eliminating fields or field components in configuration
space. 
\end{itemize}
These situations must be worked in a case by case basis.
\begin{itemize}
\item[{\it{iii)}}]
All the constraints $\chi^A$ are holonomic, hence the above 
theorem applies,
the stabilization mechanism can be run again and the three possibilities 
we are examining are open for the next step.  
Let us consider the most favorable case, i.e. the constraints are
always holonomic. 
A possible outcome may be that after a
certain number of steps, no new constraints appear. In fact if the
algorithm does not stop, since the number of d.o.f. per space-time
point is finite, we shall end up with no d.o.f. at all,
thus signaling the incompatibility of the constraints $f^A=0$ with
the dynamics derived from the original Lagrangian ${\mathcal L}$. 

On the other hand, if the algorithm stops the following proposition can be 
proven

\vspace{6mm}

\noindent {\bf Proposition:} {\it If all constraints,
$\chi^A$ and their subsequent stabilizations, are holonomic 
and non-gauge-fixing, and the algorithm stops, 
the final theory is a consistent truncation of the original one}. 

\vspace{6mm}

The proof is immediate just by examining the last step of the algorithm.
In fact this is the case in both of the examples discussed below. 
This result can also be given a geometric flavor as follows. Consider
the tangent space ${\mathcal T}Q$ of a configuration manifold $Q$, and some
dynamics defined in it by means of a variational principle. If 
$\bar Q$ is a submanifold of $Q$, such that the dynamics in ${\mathcal T}Q$
is tangent to the submanifold  ${\mathcal T}\bar Q$, then the variational 
principle can be directly formulated in ${\mathcal T}\bar Q$. 
This is just the formulation given by the reduced Lagrangian, thus providing
with a consistent truncation of the original theory.

\end{itemize}

\vspace{6mm}

Another mechanism to guarantee that a truncation driven by constraints is 
consistent is provided by symmetry considerations. One can state

\vspace{6mm}

\noindent {\bf Theorem 2:} {\sl If the constraints $f^A$ generate, via the 
Poisson brackets, a symmetry 
of the e.o.m., then the truncation is consistent.}

\vspace{6mm}

The proof is given in the appendix.

\section{Applications: general structure}
\label{appl}

In order to exemplify the above general results we shall work out
explicitly two bosonic solutions characterized by: {\it i)} both
are solutions of theories obtained as low-energy field theory
limits of type-IIB string theory and {\it ii)} the solutions
are obtained via uplifting from $7d$ to $10d$ after the
truncation of the initial theory to $7d$. As both solutions are 
related we shall make a common treatment 
of the first steps in the truncation and uplifting procedure.

Before proceeding to the subject proper, however, we must first discuss
briefly some technical details:
the space-time manifold we shall consider is constituted
by a non-compact part times a compact one, ${\mathcal M}_d \otimes
{\mathcal H}_n$. 
The requirement of compacity ensures the factorization of  a finite
volume in the action.
In the
cases we shall develop below, we identify ${\mathcal H}_n \equiv
S^3$, where the $S^3$ is considered as the group manifold $SU(2)$.
Its isometry group is $SO(4) = SU(2)\times SU(2)$ but we assume
that only one of the $SU(2)$'s is an isometry of the full metric on
${\mathcal M}_d \otimes {\mathcal H}_n$.
In particular we shall tackle two
solutions \cite{strominger,chamseddine} that can be thought as upliftings
after a previous truncation of the theory
from $10d$ to $7d$. In this case the Lie
algebra of Killing vectors, $K_{a}=
K_{a}^\alpha(y)\partial_\alpha$, with $[{K}_{a},{K}_{b}] =
C^{c}_{ab}{K}_{c}$, used in the dimensional reduction, is that of
$su(2)$. The $y^\alpha$'s denote the $S^3$ coordinates and will 
eventually disappear in the truncation.
As a matter of notation, the indices
$\alpha$ and $a$ run over $1,2,3$, and $\psi^1,\
{\theta^1},\ \phi^1$ are the parameters of the sphere $S^3$. One can
construct a basis of left-invariant one-forms, $\omega^{a}=
\omega^{a}_\alpha(y) {d} y^\alpha$,
with  ${\mathfrak L}_{{K}_a}\omega^{b}=0$, where
${\mathfrak L}_{{K}_a}$ stands for the Lie derivative with respect
to the vector ${K}_a$. 
In our conventions,
\begin{equation}
\label{ws}
\omega^1 + i \omega^2 = \frac{1}{2}
e^{-i \psi^1} \left(d{\theta}^1+i\sin{\theta}^1
d\phi^1\right)\,,\quad
\omega^3= \frac{1}{2} \left( d\psi^1+\cos\theta^1 d\phi^1\right) \,.
\end{equation}
This basis satisfies the Cartan-Maurer
equations
$
d\omega^a = \frac{1}{2} C^a_{bc}\, \omega^b\,\wedge \omega^c\,,
$
with the structure constants $C^a_{bc} = -2 \epsilon_{abc}$\,.

The manifold is also provided with a rank-2 symmetric tensor, the
metric field ${g}$, satisfying the Killing conditions, ${\mathfrak
L}_{{K}_a}{g}=0$. Written in the mixed basis ${d} x^\nu,\
{\omega}^{a}$, it takes the form
\begin{equation}
{g} =  g_{\mu\nu}(x)\, {d}x^\mu {d}x^\nu +
g_{ab}(x)\left(A^a_\mu(x)\, {d}x^\mu  +  {\omega}^{a}\right)
\left(A^b_\nu(x)\, {d} x^\nu +  {\omega}^{b}\right)\,.
\label{xmetric}
\end{equation}
Notice that we keep {\it
all} components of the metric. The
Killing conditions make them to depend only on the
$x$ variables.

After the first-type truncation the quotient manifold is parametrised by the
$x^\mu$ coordinates ($\mu$ = $0,1,\dots,6$).

\subsection{Setting the framework}
\label{tr}

{F}or the two solutions at hand
the starting point will be the bosonic
sector\footnote{This sector by its own is a
consistent truncation of the full theory.} of
type IIB supergravity ($d+n=7+3=10$) \cite{johmson}
\begin{eqnarray}
S_{\rm S}^{(d+n)} &=& \frac{1}{2 \kappa_0^2} \int  \, d^d x\,d^n
y\, \sqrt{-\hat g}\, \left\{ e^{-2 \hat\Phi}\, \left[ {\mathcal{
\hat R}} + 4\, \partial_{\hat\mu}
 \hat\Phi\,
\partial_{\hat\nu} \hat\Phi\, \hat{g}^{\hat\mu \hat\nu}
-\frac{1}{12}  \left(\hat{H}^{(3)}\right)^2 \right] \right.
\nonumber \\ && \left. -\frac{1}{12} \left( \hat{G}^{(3)} + \hat{
C}^{(0)}\, \hat{H}^{(3)} \right)^2 - \frac{1}{2} d\hat{ C}^{(0)}\,
d\hat{ C}^{(0)} -\frac{1}{480} \hat{G}^{(5)}\, \hat{G}^{(5)}
\right\} \nonumber \\ && + \frac{1}{4 \kappa_0^2} \int \left( \hat
C^{(4)} + \frac{1}{2} \hat{ B}^{(2)}\, \hat{ C}^{(2)} \right)
\hat{ G}^{(3)}\, \hat{ H}^{(3)}\,,
\end{eqnarray}
where $\hat{H}^{(3)} := d \hat{ B}^{(2)}$ is the field strength of
the NS two-form,
\begin{equation}
\left(\hat{H}^{(3)}\right)^2 := \hat{H}_{\hat\mu\hat\nu\hat\rho}\,
\hg^{\hat\mu \hat\alpha}\, \hg^{\hat\nu\hat\beta}\,
\hg^{\hat\rho\hat\gamma}\,
\hat{H}_{\hat\alpha\hat\beta\hat\gamma}\,,
\end{equation}
and $\hat{G}^{(3)} := d\hat{C}^{(2)}\,, \hat{G}^{(5)} : =
d\hat{C}^{(4)}+ \hat{ H}^{(3)} \hat{ C}^{(2)}$ are the RR field
strengths. In addition $\hat{C}^{(0)}$ stands for the RR scalar
field, the axion. 
{F}or the time being $n$ and $d$ are kept generic.
Notice that we can consistently set to zero the
RR field strengths and the RR scalar field without any further
implication in the theory. This is due to the fact that the
appearance of these fields inside the action is quadratic and
hence they identically imply a vanishing equation of motion when
set to zero. At this early stage, and after substitution of the
inverse matrix corresponding to \bref{xmetric},
\begin{eqnarray}
\label{invers}
\hat{g}^{\hat\mu \hat\nu}
 = \, \left( \begin{array}{cc}
g^{\mu \nu} & - g^{\mu\nu}\,A^a_\mu\, Y^\beta_a\\
-g^{\rho\nu}\, A^a_\rho\, Y^\alpha_a &
\left( g^{ab} + g^{\mu\nu}\, A^a_\mu\, A^b_\nu \right)
Y^\alpha_a Y^\beta_b
\end{array}  \right)\,,
\end{eqnarray}
we shall only deal with the expression between squared brackets in
\bref{iib}, i.e.
\begin{equation}
\label{iib}
S_{\rm S}^{(d+n)} = \frac{1}{2\kappa_0^2}
\int  \, d^{d} x\,d^{n} y\, \vert -\hat g_{\hat\mu\hat\nu} \vert^{1/2}\,
e^{-2 \hat\Phi}\, \left( \hat{\mathfrak{ R}} + 4\,
\partial_{\hat\mu} \hat\Phi\, \partial_{\hat\nu} \hat\Phi\,
\hat{g}^{\hat\mu \hat\nu} -\frac{1}{12} (\hat{H}^{(3)})^2\,
\right)\,,
\end{equation}
that constitutes already a consistent truncation of type IIB
theory. One can verify in addition that it also corresponds to a consistent
truncation of type I supergravity.

\subsection{Truncation from $10d$ to $7d$}

In the two examples that are examined later, the total
truncation can be performed in several steps, combining the two 
truncations, first-type and second-type, discussed before. Here we shall 
display the first two steps,
common to the two examples, exhibiting already the main features
of each type of truncation.

\subsubsection{First step: pure dimensional reduction}
The theory is
truncated by a pure dimensional reduction from $(d+n)$ dimensions
(coordinates $x,\,y$) to 
$d$ dimensions
(coordinates $x$), keeping unchanged the number of
d.o.f.  per space-time point
$${\mathcal L}^{(d+n)}(\hat g_{\hat\mu\hat\nu} ,\
\hat\Phi,\  \hat B_{\hat\mu\hat\nu})\longrightarrow {\mathcal
L}_{\!\rm R}^{(d)}( g_{\mu\nu},\ g_{ab},\ A^a_\mu,\ \Phi,\
B_{\mu\nu},B_{\mu a},B_{ab})\,.$$ The associated Lagrangian density, $
{\mathcal L}_{\!\rm R}^{(d)}$, is a first-type consistent truncation
of $ {\mathcal L}^{(d+n)}$ with an expression which deduced from
the $(d+n)$-dimensional Einstein frame takes the form
\cite{Scherk:1979zr,chofreund}
\begin{eqnarray}
\label{lr}
{\mathcal L}^{(d)}_{\rm R}& =& \frac{{\rm Vol_n}}{2\kappa_0^2}\, \vert -
g_{\mu\nu}\vert^{1/2} \vert g_{ab}\vert^{1/2} \Bigg\{ {\mathfrak
R}- \frac{1}{4} F^{\mu\nu a}\, F_{\mu\nu}^b\, g_{ab}\nonumber
+\frac{1}{4} g^{\mu\nu}\, {\mathcal D}_\mu g_{ab}\, {\mathcal
D}_\nu g^{ab}
\nonumber \\&&
+g^{\mu\nu}\, {\partial}_\mu \ln \sqrt{g_{ab}} \,
{\partial}_\nu \ln \sqrt{g_{ab}}
-\frac{1}{4}\,C^a_{bc}\left[ 2C^b_{a c^\prime} \, g^{c c^\prime} +
C^{a^\prime}_{b^\prime c^\prime}\, g_{aa^\prime}\, g^{bb^\prime}
\,g^{cc^\prime} \right] \nonumber \\&&
-\frac{1}{2} \left( \nabla \Phi \right)^2 -\frac{1}{12} e^{-\Phi}
\left( H^{(3)}\right)^2
\Bigg\} \,, \label{ss-formula}
\end{eqnarray}
being ${\mathcal D}_{\!\mu}$ the covariant derivative for the
Yang-Mills connection and ${\rm Vol_n}=\int \vert \omega\vert d^n y$.
It is not necessary to express the
components of the NS-NS gauge field strength.
The legitimacy of the procedure
is guaranteed by the tracelessness condition, $C^a_{ab} = 0$.
Any solution of the e.o.m.\! for ${\mathcal L}_{\!\rm
R}^{(d)}$ can be uplifted to a solution of the e.o.m.\! for ${\mathcal
L}^{(d+n)}$, and vice-versa, any solution of the e.o.m.\! for
${\mathcal L}^{(d+n)}$ that satisfies the Killing conditions can
be obtained as an uplifting of a solution of the e.o.m.\! for
${\mathcal L}_{\!\rm R}^{(d)}$.

\avall

Hitherto the space-time dimension has been general, henceforth, and
for sake of clarity, we set $d=7, n=3$. The Lie group of Killing
symmetries is $SU(2)$.

\subsubsection{Second step: Truncation driven by constraints}
It
consists in truncating the scalar spectrum of the theory. This is
a second-type truncation, driven by constraints, and its
consistency relies in the findings of sec.~\ref{general}. We introduce,
in the ${\mathcal L}_{\!\rm R}^{(7)}$ theory, the
constraints\footnote{To introduce some constraints in a theory
means that we look for solutions of the e.o.m.\! of the theory
that in addition satisfy such constraints.}
\begin{equation} g_{ab}C^a_{cd} +
g_{ad}C^a_{cb} = 0 \,. \label{simple-case}
\end{equation}
For a simple Lie algebra, where with
our conventions the Cartan-Killing constant metric $h_{ab} :=
C^c_{ad}C^d_{bc}$ is a multiple of $\delta_{ab}$, equation
\bref{simple-case} is equivalent to the more familiar relation
\begin{equation}  g_{ab}=
\varphi \delta_{ab}\,, \label{simple-case2}
\end{equation}
for some $\varphi$ that becomes the only remaining scalar field
from those originated from the $10$-dimensional metric. These
constraints eliminate all the scalars that carry YM charges and
make the internal manifold bi-invariant, under two copies of the
simple Lie group. 
We have
$${\mathcal L}_{\!\rm R}^{{(7)}}( g_{\mu\nu},\
g_{ab},\ A^a_\mu,\ \Phi,\
B_{\mu\nu},B_{\mu a},B_{ab})\longrightarrow{\mathcal L}_{\!2\rm R}^{{(7)}}(
g_{\mu\nu},\ \varphi,\ A^a_\mu,\ \Phi,\ B_{\mu\nu},B_{\mu a},B_{ab})\,,$$
with the r.h.s. given by
\begin{eqnarray}
\label{l2reducedmn} {\mathcal L}_{\!2 \rm R}^{{(7)}}& =&
\frac{{\rm Vol_3}}{2\kappa_0^2}\, \vert - g_{\mu\nu}\vert^{1/2}\,
\varphi^{3/2} \left( \mathfrak{R} - \frac{1}{4} \varphi F^a F^b
\delta_{ab} +\frac{3}{2}  \left(\frac{\nabla
\varphi}{\varphi}\right)^2 +\frac{6}{\varphi} - \frac{1}{2}
\left(\nabla\Phi\right)^2 
\right. \nonumber \\&& \left.
-\frac{1}{12} e^{-\Phi} \left(\hat{H}^{(3)} \right)^2\Big|_{\!
g_{ab} = \varphi\delta_{ab}} \right)\,.
\end{eqnarray}

As regards the e.o.m.\! it turns out that
\begin{eqnarray}
\label{eoms} \left(\frac{\delta {\mathcal L}_{\rm
R}^{{(7)}}}{\delta g_{\mu\nu}}\right)_{\!g_{ab}= \varphi
\delta_{ab}} &=& \frac{\delta {\mathcal L}_{2 \rm
R}^{{(7)}}}{\delta g_{\mu\nu}}\,, \nonumber \\ \label{second-red}
\left(\frac{\delta {\mathcal L}_{\rm R}^{{(7)}}}{\delta
A^a_{\mu}}\right)_{\!g_{ab}= \varphi \delta_{ab}} &=& \frac{\delta
{\mathcal L}_{2 \rm R}^{{(7)}}}{\delta A^a_{\mu}}\,,\\
\left(\frac{\delta {\mathcal L}_{\rm R}^{{(7)}}}{\delta
g_{ab}}\right)_{\!g_{ab}= \varphi \delta_{ab}} &=&
\frac{1}{3}\left(\frac{\delta {\mathcal L}_{2 \rm
R}^{{(7)}}}{\delta \varphi}\right)\delta^{ab} - \left(\frac{{\rm Vol_3}}{2
\kappa_0^2}\,{\vert{g_{\mu\nu}}\vert}^{\frac{1}{2}}
\varphi^{\frac{3}{2}}\right) \chi^{ab} \,, \nonumber
\end{eqnarray}
with
$$
\chi^{ab} := \frac{1}{4}M^{ab}- \frac{1}{12}(M^{cd}\, \delta_{cd})
\delta^{ab}\,,\quad
M^{ab} := F^a_{\mu\nu} F^{\mu\nu b} -
\frac{e^{-\Phi}}{{\varphi}^{2}} \, H_{c\hat\mu\hat\nu}\, H^{\ \,
\hat\mu\hat\nu}_d\, \delta^{ca}\, \delta^{db}\,.
$$
At first sight we realize that the truncation is inconsistent
because the presence of a term proportional to $\chi^{ab}$ in the
r.h.s. \bref{second-red}, which may be different from zero. What
we have found are just the secondary constraints predicted by our
theorem of sec.~\ref{general}. {F}or a simplified model with the same
features see \cite{pere-pep}. We must therefore require
\begin{equation}
\label{chi}
\chi^{ab} = 0\,,
\end{equation}
on the candidate configurations for an uplifting from the
${\mathcal L}_{\!2 \rm R}^{(7)}$ theory to the ${\mathcal L}_{\! \rm
R}^{(7)}$. 

\section{Applications: specific models}
\label{appl2}

After showing the general pattern to be followed, we find it
worthwhile to work out two specific examples that comply the
requirement of being built on semisimple Lie algebras. As we
mentioned already in sec.~\ref{appl} both examples consider
basically a truncation of $10d$ supergravity to $7d$ with a
different content of supersymmetry.

\subsection{IIB NS-$5$ branes}
Let us consider the consistent bosonic truncation of Type IIB,
which in the Einstein frame takes the form
\begin{equation}
\label{xc} S_{\rm E}^{(10)} = \frac{1}{2\kappa_0^2} \int  \, d^{10}
x\, \vert -\hat g_{\hat\mu\hat\nu} \vert^{1/2}\, \, \left(
\hat{\mathfrak{ R}} -\frac{1}{2}\,
\partial_{\hat\mu} \hat\Phi\, \partial_{\hat\nu} \hat\Phi\,
\hat{g}^{\hat\mu \hat\nu} -\frac{1}{12} e^{-
\hat\Phi}(\hat{H}^{(3)})^2\, \right)\,.
\end{equation}
The reduced
Lagrangian can be read of from \bref{lr}.

As is analyzed in \cite{pere-pep}, and in agreement with the
theorem formulated in section sec.~\ref{general}, the consequence of imposing
the constraints \bref{simple-case} is the emergence of new,
secondary constraints, that take the form \bref{chi}.
A strong way to satisfy these new constraints is to impose on
the theory a new, drastic second-type truncation:
$$
A_\mu^a = H_{\mu\nu\sigma}=H_{\mu\nu a}=H_{\mu ab}=0\,,
$$
which is trivially consistent because the Lagrangian is quadratic
in these field components that are set to zero.

After imposition of this new truncation we obtain a newly reduced
Lagrangian which is already a consistent truncation of Type IIB
supergravity, with no constraints attached,
\begin{eqnarray}
 {\mathcal L}_{\!3 \rm R}^{(7)}& =&
\frac{{\rm Vol_3}}{2\kappa_0^2}\, \vert - g_{\mu\nu}\vert^{1/2}\,
\varphi^{3/2} \left( \mathfrak{R} +\frac{3}{2} \left(\frac{\nabla
\varphi}{\varphi}\right)^2 +\frac{6}{\varphi} - \frac{1}{2}
\left(\nabla\Phi\right)^2 
-\frac{1}{12} e^{-\Phi} \varphi^{-3}\left(H \right)^2 \right)\,,
\end{eqnarray}
where $(H)^2
=H_{abc}\delta^{aa'}\delta^{bb'}\delta^{cc'}H_{a'b'c'}$.

Moving to the Einstein frame, and defining $\varphi = e^{\Psi}$,
we re-express this theory as
\begin{eqnarray}
 {\mathcal L}_{ \rm E}^{(7)}& =&
\frac{{\rm Vol_3}}{2\kappa_0^2}\, \vert - g_{\mu\nu}\vert^{1/2}\, \left(
\mathfrak{R} -\frac{6}{5} \left(\nabla \Psi\right)^2 +6
e^{-\frac{8}{5}\Psi}- \frac{1}{2} \left(\nabla\Phi\right)^2
-\frac{1}{12} e^{-\Phi}e^{-\frac{18}{5}\Psi}
\left(H \right)^2 \right)\,.
\end{eqnarray}

Notice that the object $H_{abc}$, completely antisymmetric, is
just the product of a $7d$ scalar times $\epsilon_{abc}$. But
since, when uplifted to $10d$, it is interpreted as giving rise to
the components of a three-from field strength, $H =
H_{abc}\,\omega^a\wedge\omega^b\wedge\omega^c$, in order for $H$
to be a closed form satisfying the Killing conditions, this scalar
must be simply a constant.

This Lagrangian can undergo a new second-type truncation by
linking the two scalars, $\Phi$ and $\Psi$. Let us consider the
new constraint
$$
\chi := \Psi + \lambda \Phi =0\,,
$$
where the constant parameter $\lambda$ will be determined below by
requiring the truncation to be eventually consistent. The e.o.m.
for $\Phi$ and $\Psi$ are
\begin{eqnarray} \triangle \Phi +
\frac{1}{12}e^{-\Phi}e^{-\frac{18}{5}\Psi}(H)^2
&=&0\,, \nonumber \\
\triangle \Psi + \frac{1}{8}e^{-\Phi}e^{-\frac{18}{5}\Psi}(H)^2 -
4e^{-\frac{8}{5}\Psi}&=&0\nonumber \,,
\end{eqnarray}
and therefore
$$
\triangle \chi + (\frac{3+2\lambda}{24})
e^{-\Phi}e^{-\frac{18}{5}\Psi}(H)^2 - 4e^{-\frac{8}{5}\Psi}=0\,.
$$
Since the constraint $\chi$ is set to zero, the secondary
constraint is isolated as
$$
(3+2\lambda)e^{-\Phi-2\Psi}(H)^2= 96\,.
$$
We can make the choice $\lambda =\frac{1}{2}$. This makes the
exponent in the last expression $-\frac{1}{2}\chi$, which is zero
because we are implementing $\chi=0$. We end up with the new
constraint
$$
(H)^2= 24\,,
$$
or, equivalently,
\begin{equation} H_{abc} = \pm 2
\epsilon_{abc}\label{lastconstraint}\,.
\end{equation}
This last constraint is in fact very adequate, because we have
argued before that the scalar in $H_{abc}$ must be a constant.
Note that any other choice, $\lambda \neq\frac{1}{2}$, would have
led to inconsistencies.

One can check that the new constraint \bref{lastconstraint} is
compatible with the dynamics without the appearance of new
constraints. We have therefore proved that implementing both
constraints, $\Psi =-\frac{1}{2}\Phi$ and $H_{abc} = \pm 2
\epsilon_{abc}$, directly into the Lagrangian makes the new
truncation still consistent. The final Lagrangian is
\begin{equation}
\label{family0} {\mathcal L}_{ \rm E}^{(7)}= \frac{{\rm Vol_3}}{2\kappa_0^2}\,
\vert - g_{\mu\nu}\vert^{1/2}\, \left( \mathfrak{R} -\frac{4}{5}
\left(\nabla\Phi\right)^2  +4 e^{\frac{4}{5}\Phi} \right)\,,
\end{equation}
and our construction allows us to assert that this Lagrangian is a
consistent truncation of Type IIB supergravity. Any solution of
\bref{family0} can be uplifted to a solution of Type IIB. Notice
that in the process of uplifting from $7d$ to $10d$ a field
strength $H =
2\epsilon_{abc}\,\omega^a\wedge\omega^b\wedge\omega^c$ will
appear, together with the required components for the $10d$
metric.

The Lagrangian \bref{family0} belongs to an interesting family of
Lagrangians, one for each space-time dimension $d>2$, with a
metric and a scalar field,
\begin{equation} \label{family} {\mathcal L}_{ \rm E}^{(d)}= 
{\rm const.}\,
\vert - g_{\mu\nu}\vert^{1/2}\, \left( \mathfrak{R} -\frac{4}{d-2}
\left(\nabla\Phi\right)^2  + 4 e^{\frac{4}{d-2}\Phi} \right)\,,
\end{equation}
which are related among themselves by toroidal (i.e., Abelian)
consistent truncations. In fact, every single truncation includes
{\it i)} a one-dimensional reduction, {\it ii)} a second-type truncation to get
rid of a Maxwell field, {\it iii)} another second-type truncation to link
the scalar mode coming from the reduction of the metric with the
dilaton, and {\it iv)} a conformal redefinition of the metric in order to
stay in the Einstein frame. It is easy to verify that the
following conformally flat configuration
\begin{eqnarray}
g^{(d)}_{\mu\nu} &=& e^{\frac{4}{d-2}\rho}\ \eta^{(d)}_{\mu\nu}\,,
\nonumber \\\Phi &=& -\rho\,,
\end{eqnarray}
(where $\eta^{(d)}_{\mu\nu}$ is the $d$-dimensional Minkowski
metric and $\rho$ is a spatial-coordinate) is a solution of the
theory \footnote{ The factor of $4$ in the potential term in \bref{family}
can be generalized to an arbitrary $\beta > 0$. Then $\Phi$
changes to $\Phi = -\frac{\sqrt{\beta}}{2}\rho$.}. In fact {\sl
these solutions for different dimensions are all connected by
upliftings}. When uplifted from $7d$ to $10d$ by use of the
$su(2)$ Killing algebra we obtain a solution of Type IIB that, in
Einstein frame, takes the form
\begin{eqnarray}
\left(ds^{(10)}\right)^2 &=& e^{\frac{1}{2}\rho}\left( ds^2\left(\mathbb{
E}^{(1,6)}\right)+d\rho^2 +\sum_{a=1}^3\omega^a \otimes \omega^a
\right)\,, \nonumber \\ \Phi &=& -\rho\,, \nonumber \\ H &=&
2\epsilon_{abc}\,\omega^a\wedge\omega^b\wedge\omega^c\,.
\end{eqnarray}
This is the well known solution describing an extremal
Dp-brane \cite{strominger}.

\subsection{D5-branes wrapping $S^2$}
\label{mns}

As a second exemplification of the theorem we present a much less
straightforward model, 
which also describes a solution of \bref{xc}. This model describes 5-branes
wrapping a two cycle and is dual in the infrared to   
${\mathcal N} = 1$ super Yang-Mills  \cite{mn}. It has also been constructed
without relying on upliftings in \cite{arkady2,gubser,arkady}. 
The first-type truncation follows the same lines as in the previous
example, thus the model under consideration will result from uplifting a
solution of \bref{lr}.
It contains a nonzero NS-NS two-form field given generically by
\begin{equation}
\label{h3}
H^{(3)}=dB^{(2)} = 2 \left(\omega^1+A^1\right)\wedge
\left(\omega^2+A^2\right)\wedge \left(\omega^3+A^3\right) +
\sum_{a=1}^3 F^a \wedge \left(\omega^a+A^a\right)\,,
\end{equation}
where $\omega_i$ are given in \bref{ws} and the components of the 
$SU(2)$ YM potential $A^a$ are \footnote{
We shall share the definitions of \cite{paolo} with a
value of their parameter $\lambda = -1$.}
\begin{equation}
\label{as} A^{1}= \frac{1}{2}  a(\rho) d\theta\,, \quad A^{2}= -
\frac{1}{2}  a(\rho) \sin\theta d\varphi\,, \quad
A^{3}=\frac{1}{2}  \cos\theta d\varphi\,.
\end{equation}

The complete display of the model includes 
the metric, which in the string frame reads
\begin{equation}
\label{metricmn} \left(ds^{(10)}\right)^2 =  ds^2\left(\mathbb{ E}^{(1,3)}\right)
+  d\rho^2 + e^{2g(\rho)} d\Omega^2_2+ \sum_{a=1}^3
\left(\omega^a+A^a\right)^2\,,
\end{equation}
and the dilaton field, 
\begin{equation}
\label{dil-NS5} e^{2\hat{\phi}(\rho)}=
\frac{2e^{g(\rho)}}{\sinh2\rho}\,.
\end{equation}
The functions $a(\rho)$ and $g(\rho)$ remain to be determined. These
were first obtained in a $7d$ context in
\cite{chamseddine,cv},
\begin{equation}
\label{lastdil}
a(\rho):= \frac{2\rho}{\sinh 2\rho}\,,
\end{equation}
and
\begin{equation}
\label{lastmn} e^{2g(\rho)} := \rho \coth 2\rho -
\frac{\rho^2}{\sinh^2 2\rho} -\frac{1}{4}\,.
\end{equation}
Notice that this last function is highly nontrivial and it
\emph{runs} on the $\rho$ direction; this fact invalidates the
premises of \cite{runing}, and hence allows \emph{in principle}
for a dimensional reduction together with the elimination of
degrees of freedom. 
\avall

Let us summarize the present status. We have on one hand   
a solution of a $7d$ supergravity. In addition, there is a set 
of constraints, \bref{chi}, that, according to the theorem in 
sec.~\ref{general}, play the 
role of conditions for the uplifting. That is, 
the $7d$ solution will be upliftable to a solution of the $10d$ theory
if, and only if, it satisfies \bref{chi}.  

The satisfaction of \bref{chi} is guaranteed by the structure 
of \bref{h3}, independently of the explicit form of $a(\rho)$ and
$g(\rho)$, provided that $\varphi^2 = e^{-\Phi}$ in the 10d
Einstein frame --corresponding to $\varphi = 1$ in the string frame--.
As a consequence, the 7d solution of
\cite{chamseddine,cv}, which has been already displayed in the string frame,
is upliftable to 10d \emph{if}  
the constraint $\varphi=1$
is fulfilled. This condition is indeed satisfied by \bref{metricmn}, and 
therefore we conclude that the $7d$ solution (\ref{h3}--\ref{lastmn}) 
is upliftable to a solution of Type IIB.

Let us mention that, as in the previous example, 
if one starts with the right set of constraints \cite{Cvetic:2003jy}, 
the reduction algorithm can be applied several times, as described in 
sec.~\ref{general}. In this case one can end up with a consistent truncation 
of type IIB which still exhibits a non-trivial YM gauge potential 
\cite{Cvetic:2003jy,Duff:1985cm}.

\section{Conclusions}
\label{conclusions}

Despite the big amount of recent literature concerning the
truncation and uplifting procedures we find compelling to revise in a original
way some models that are commonly used in the
context, or in generalizations, of the AdS/CFT duality. 
In this respect we formulate in a systematic manner the necessary and sufficient conditions, 
for the construction of supergravity models via uplifting procedures. 
{F}or this purpose we provide with some theoretical results that stablish under
which conditions a truncation of a theory, with holonomic constraints, 
can lead to consistent upliftable solutions.

We disentangle in a neat fashion two different steps in
the truncation procedure. A first one deals solely with dimensional reduction 
and a second with the elimination of d.o.f. in configuration space.
While the former has been considered elsewhere \cite{pere-pep} 
in the case of quotienting out a group manifold,
we have concentrated here in elucidating the
consequences of the latter. Is it clear that the first and
unavoidable consequence of eliminating some d.o.f. is the existence of
classical constraints that have to be satisfied. The theorem in sec. 
\ref{general} establishes that the 
theory defined with the remaining d.o.f. must in general be supplemented with secondary
constraints. If that is the case the truncation is not consistent by its own. Anyhow
correct upliftings can be obtained as long as we consider
solutions of the reduced theory that satisfy these \emph{new} constraints.
In the case that the new constraints are defined in configuration space the 
truncation procedure can be applied again. In this way it is possible to end up with
a consistent truncation when no more constraints appear at a given stage.

Our approach have been applied in two models. The first one for its
simplicity is just a plain illustration of the different steps,
while the second is the only known non-conformal, ${\mathcal N}=1$
SYM supergravity solution.

\vskip 6mm

{\it{\bf Acknowledgments}}

We thank Carlos N\'u\~nez for suggestions and discussions and
Arkady Tseytlin for criticism. The work of J.\ M.\ P.\ is
partially supported by MCYT FPA, 2001-3598, CIRIT, GC
2001SGR-00065, and HPRN-CT-2000-00131.

\appendix
\setcounter{equation}{0}
\newcounter{zahler}
\addtocounter{zahler}{1}
\renewcommand{\thesection}{\Alph{zahler}}
\renewcommand{\theequation}{\Alph{zahler}.\arabic{equation}}
\label{appdx}

\setcounter{section}{0}
\setcounter{subsection}{0}

\renewcommand{\thesection}{\Alph{zahler}}
\renewcommand{\theequation}{\Alph{zahler}.\arabic{equation}}

\section{Proof of the theorems}
\label{proof}

Second type truncations are characterized by the elimination of degrees of freedom per
space-time point. Here we shall consider the 
case when the constraints that reduce the degrees of freedom are
holonomic, that is, when they constrain the variables, fields, in
configuration space.

\subsection{Notation and preliminaries}

{F}or the sake of simplicity in the exposition, we shall use in this section the language of
mechanics. The analogous results for field theories are straightforwardly obtained by
making use of De Witt's compact notation \cite{dewitt}, in which the labels for our objects can
represent not only discrete but continuous indices as well.
As a matter of notation the evolutionary parameter --the time-- is denoted by
$t$, and the configuration-space variables are $q^i$. With the appropriate change of
variables we can explicitate the constraints as some of the coordinates. This means
that we can take $q^i= (q^A, q^a)$ with the constraints represented by $q^A =0$. The
remaining variables $q^a$ are the variables for the reduced configuration space.

Consider a Lagrangian $ L(q^i,\dot q^i)$. 
Plugging the constraints $q^A =0,\, \dot q^A =0$,
into it yields
the reduced Lagrangian $L_{\rm R}(q^a,\dot q^a) 
= L(q^A=0,q^a ,\dot q^A=0,\dot q^a)$.
The original e.o.m. are
\begin{equation}
[ L]_i = \alpha_i - W_{ij}\ddot q^j = 0\,,
\label{el-eq}
\end{equation}
where
\begin{equation} 
W_{ij}:= {\partial^2 L\over\partial\dot
q^i\partial\dot q^j},
\label{hess}
\end{equation}
is the Hessian matrix for $ L$ w.r.t. the velocities
and 
$
\alpha_i :={\partial L\over\partial q^i}
    - {\partial^2  L\over\partial\dot q^i\partial q^j}\dot q^j
     \,.
$

When the reduction, $q^A =0, \dot q^A =0, \ddot q^A =0$, is implemented at the level
of the e.o.m. we get, for $i=a$,
$$
\left([ L]_a\right)_{\rm R} =
(\alpha_a)_{\rm R} - (W_{ab})_{\rm R}\, \ddot q^b = \tilde\alpha_a- \tilde W_{ab}\,\ddot q^b =
[ L_{\rm R}]_a = 0\,,
$$
where
$$\tilde\alpha_a:={\partial L_{\rm R}\over\partial q^a}
    - {\partial^2 L_{\rm R}\over\partial\dot q^a\partial q^b}\dot q^b\,,
$$
and $\tilde W_{ab}$ is the Hessian matrix for $ L_{\rm R}$ 
(tildes 
quantities will correspond to the reduced formalism, derived from $ L_{\rm R}$).
In addition, for $i=A$ 
$$
\left([ L]_A\right)_{\rm R} 
= (\alpha_A)_{\rm R} - (W_{Ab})_{\!{\rm R}}\,\ddot q^b=0\,.
$$

An immediate consequence is that this kind of reduction will generally be inconsistent
because the reduced Lagrangian only produces a part of the reduced e.o.m.
$$
\left\{
\begin{array}{c}
[ L]_i =0 \\ q^A=0
\end{array} \right\} \ \Longleftrightarrow \ ([ L]_i)_{\rm R} 
=0 \ \ \Longleftrightarrow
\left\{ 
\begin{array}{c}
  [ L_{\rm R}]_a = 0 \\ ([ L]_A)_{\rm R}
 = 0 \end{array} \right\}
$$
Notice that the remaining part,  $([L]_A)_{\rm R} = 0$, which is not generated
by $L_{\rm R}$, seems to contain in general second derivatives of the variables. But,
as we shall see this can be avoided.
It is obviously so for theories defined with Lagrangians that are regular, 
that is,
when their Hessian matrix is regular. In such case one could have isolated 
$\ddot q^A$
from the e.o.m.  $[L]_i =0$, that is, $\ddot q^A = (W^{-1})^{Ai}\alpha_i$, and then
implement
$q^A =0\,, \dot q^A =0\,, \ddot q^A =0$ on this last relation. This gives
$$(W^{-1}\alpha)^A_{\rm R} := \big((W^{-1})^{Ai}\alpha_i\big)_{\rm R} = 0\,,$$ as new constraints,
depending only on $q^a, \dot q^a$. Therefore, for theories defined with regular
Lagrangians, the following equivalence holds
$$
([L]_i)_{\rm R} =0 \ \ \Longleftrightarrow \left\{\begin{array}{c}
[L_{\rm R}]_a = 0
\\ (W^{-1}\alpha)^{A}_{\rm R} = 0 \end{array} \right\}
$$

Our cases of interest, though, are theories allowing for gauge freedom, 
and this implies
that the Hessian matrix of our Lagrangians must be singular. In such case,
the constraints
introduced by way of the second-type truncation can be classified according to as
to whether
they restrict or not the gauge freedom. 
We shall address the case when the truncation constraints do not restrict the gauge freedom.
In order to express this condition, it is convenient to write the original e.o.m. in an
equivalent form. To do so one must consider a basis, $\gamma_\mu^i$, of the null
vectors of the Hessian matrix, that satisfy, by definition,
$$
\gamma_\mu^i W_{ij} = 0\,.
$$
It is obvious that the solutions of the e.o.m. \bref{el-eq} must satisfy
the primary constraints
$$
\alpha \gamma_\mu := \alpha_i \gamma_\mu^i = 0\,.
$$

Next consider that although the matrix $W_{ij}$ is not invertible, there exist
--non-unique--
objects $M^{ij}$ (symmetric) and $\sigma_i^\mu$ such that
\begin{equation}
W_{kj}M^{ji} + \gamma_\mu^i\sigma_k^\mu = \delta^i_k \,.\label{invert}
\end{equation}


We can express the dynamical vector field ${\bf X}$ that generates the solution
trajectories out
of some initial conditions. Saturating \bref{invert} with $\alpha_i$, and taking into
account the primary constraints, we end up with
\begin{equation}
W_{kj}(M^{ji}\alpha_i) - \alpha_k \approx 0\,,\label{inverting}
\end{equation}
where we have borrowed Dirac's notation of weak equalities for those equalities that are
satisfied on the constraints' surface. 

Comparison of \bref{inverting} with \bref{el-eq}
allows to isolate $\ddot q^j$ with its inherent gauge ambiguity:
$$
\ddot q^j = M^{ji}\alpha_i + \eta^\mu \gamma_\mu^j\,,
$$
where $\eta^\mu$ are a set of arbitrary functions that essentially reflect the
existence of gauge freedom in the theory.

Therefore the dynamical vector field is expressed as
$$
{\bf X} := \dot q^i \frac{\partial }{\partial q^i} +
(\alpha M)^i \frac{\partial }{\partial \dot q^i} + \eta^\mu \Gamma_\mu
$$
where $(\alpha M)^i := \alpha_k M^{ki}$ and \ $\Gamma_\mu = \gamma_\mu\frac{\partial }
{\partial \dot q^i}\,.$

It can be proven \cite{Batlle:1985ss} that the e.o.m. $[L] = 0$ is equivalent
to the assertion that the dynamics is generated by ${\bf X}$ on the
primary constraints' surface
\begin{equation}
[L] = 0  \quad \Longleftrightarrow  \quad \left\{
\begin{array}{c} {\bf X}\\ \alpha \gamma_\mu=0 \end{array} \right\} 
\label{equiv1}
\end{equation}

\subsection{Proof of theorem 1}

With this preparation, the assumption that the
truncation constraints, $q^A=0$, do not reduce the gauge freedom can be made
more precise. One clear way to guarantee it is by assuming 
\begin{equation}
(\Gamma_\mu \dot q^A)_{\rm R} = 0\,,\label{newas}
\end{equation}
because the requirement that these truncation constraints are preserved by the
dynamics, $({\bf X}\,\dot q^A)_{\rm R} =0$, will fix none of the arbitrary functions
$\eta^\mu$.

\avall

Three main consequences can be drawn out of \bref{newas}

\begin{itemize}
\item[{\it{i)}}]
Noticing that \bref{newas} is just $(\gamma_\mu^A)_{\rm R}=0$, one can deduce that
$(\gamma_\mu^a)_{\rm R} = \tilde\gamma_\mu^{a}$, that is, they form 
a basis for the null vectors of the Hessian matrix of the
reduced theory. On the other hand, since
$(\alpha_a)_{\rm R} =  \tilde\alpha_a$, it follows that
$$
(\alpha \gamma_\mu)_{\rm R} = \tilde\alpha \tilde\gamma_\mu\,,
$$
are just the primary constraints of the reduced theory.
\item[{\it{ii)}}]
Noting that
$$
({\bf X}\dot q^A)_{\rm R} = \big((\alpha M)^A\big)_{\rm R}\,,
$$
we infer that the time preservation of the constraints $q^A=0$ requires 
the fulfillment of
\begin{equation}
\chi^{A} :=\big((\alpha M)^A\big)_{\rm R}=0\,. \label{newcons}
\end{equation}
\item[{\it{iii)}}]
Once the new constraints \bref{newcons}
are taken into account, the vector field ${\bf X}$ reduces to
\begin{equation}
({\bf X})_{\rm R} = \dot q^a \frac{\partial }{\partial q^a} +
((\alpha M)^a)_{\rm R} \frac{\partial }{\partial \dot q^a} + \eta^\mu \tilde\Gamma_\mu\,,
\label{Xreduced}
\end{equation}
where  $ \tilde\Gamma_\mu = 
\tilde\gamma_\mu^a\frac{\partial }{\partial \dot q^a}\,.$
\end{itemize}

\avall

With the use of {\it ii)} and {\it iii)} we have the equivalence
\begin{eqnarray}
\left\{ \begin{array}{c}
{\bf X}  
\\  q^A = 0\end{array} \right\}
 \Longleftrightarrow  
\left\{ \begin{array}{c}
({\bf X})_{\rm R}\\ \chi^A=0\end{array}\right\} 
\label{equiv2}
\end{eqnarray}
which, considering \bref{equiv1} and {\it i)}, 
may also be written as 
\begin{eqnarray}([L]_i)_{\rm R} =0
\Longleftrightarrow  
\left\{ \begin{array}{c}
{\bf X}\\  \alpha\gamma_\mu = 0 \\
\ q^A = 0 \end{array}\right\}
 \Longleftrightarrow  \left\{ \begin{array}{c}
({\bf X})_{\rm R} \\
\tilde\alpha\tilde\gamma_\mu = 0 \\ \chi^A=0\end{array}\right\}
\label{equiv20}
\end{eqnarray}

Let us look closely at the piece $((\alpha M)^a)_{\rm R}$ in \bref{Xreduced}. Saturating again
\bref{invert} with $\alpha_i$ and then reducing the result to $q^A = 0, \dot q^A = 0$,
 one gets, in particular,
$$
(W_{ab})_{\rm R}(M\alpha)^b_{\rm R} + (W_{aB})_{\rm R}(M\alpha)^B_{\rm R} =  (\alpha_a)_{\rm R} = \tilde \alpha_a
$$
but $(W_{ab})_{\rm R}$ is just the Hessian of the reduced Lagrangian $L_{\rm R}$,
$(W_{ab})_{\rm R} = \tilde W_{ab}$, and  $(M\alpha)^B_{\rm R}$ have been already identified as new
constraints, $(M\alpha)^B_{\rm R}\approx 0$. All in all we have
$$
\tilde W_{ab}(M\alpha)^b_{\rm R} \approx \tilde \alpha_a\,,
$$
which, recalling \bref{inverting} and comparing with the e.o.m. for the reduced theory,
\begin{equation}
[L_{\rm R}]_a = \tilde\alpha_a - \tilde W_{ab}\ddot q^b = 0\,,
\label{el-eq-red}
\end{equation}
is telling us that
$$
\ddot q^b = (M\alpha)^b_{\rm R} + \eta^\mu \tilde \gamma_\mu^b\,,
$$
provided the constraints \bref{newcons} are satisfied
($\eta^\mu$ are arbitrary functions that in principle could be different from the ones
used before, but they turn out to be the same).
Therefore the dynamical vector field of the reduced theory can be expressed as
\begin{equation}
\tilde {\bf X} =  \dot q^a \frac{\partial }{\partial q^a} +
((\alpha M)^a)_{\rm R} \frac{\partial }{\partial \dot q^a} + \eta^\mu \tilde\Gamma_\mu\,,
\label{Xtilde}
\end{equation}
which coincides with $({\bf X})_{\rm R}$ in \bref{Xreduced}. Then we have obtained the
equivalence
\begin{eqnarray}
\left\{ \begin{array}{c}
({\bf X})_{\rm R}  \\ \chi^A=0 \end{array}\right\} 
 \Longleftrightarrow  \left\{ \begin{array}{c}
\tilde{\bf X}\\ \chi^A=0 \end{array}\right\}
\label{equiv3}
\end{eqnarray}

Altogether, equivalences \bref{equiv2} and \bref{equiv3} are summarized in
\begin{eqnarray}
\left\{ \begin{array}{c}
{\bf X}  \\ q^A=0 \end{array}\right\}
 \Longleftrightarrow  \left\{ \begin{array}{c}
\tilde{\bf X}\\ \chi^A=0 \end{array}\right\}
\label{equiv4}
\end{eqnarray}

But on the other hand, in a way parallel to \bref{equiv1}, the reduced Lagrangian
exhibits the equivalence
\begin{equation}
[L_{\rm R}]_a = 0   \Longleftrightarrow  
\left\{ \begin{array}{c}
\tilde{\bf X}\\  \tilde\alpha \tilde\gamma_\mu=0
\end{array}\right\}
\label{equiv5}
\end{equation}
and using both equivalences \bref{equiv4} and \bref{equiv5} together with \bref{equiv20}
we get our final result
\begin{equation}
\left\{ \begin{array}{c}
[L]_i = 0\\  q^A=0 \end{array}\right\} \Longleftrightarrow  
\left\{ \begin{array}{c} [L_{\rm R}]_a = 0\\
\chi^A=0\end{array}\right\}
\label{equiv6}
\end{equation}

\subsubsection{Final remarks}
We shall show that the assumption \bref{newas} is nothing but \bref{assumpt}. For
the sake of simplicity we have worked with coordinates $q^i$ such that the 
holonomic constraints are written as a subset $q^A=0$. In a general setting the 
constraints will be expressed in implicit form as a set of functions
$$
f^A(q) =0\,,
$$ 
and \bref{newas} will have been written as 
\begin{equation}
(\Gamma_\mu \dot f^A)_{\rm R} = 0\,,\label{newas1}
\end{equation}
with
$$
\dot f^A = \frac{\partial f^A}{\partial q^i}\dot q^i\,.
$$
On the other hand it is well known \cite{Batlle:1985ss} that a basis for the null
vectors of the Hessian matrix used to define $\Gamma_\mu$ 
is provided by the gradient of the Hamiltonian primary 
constraints $\phi_\mu$ with respect to the momenta,
$$
\gamma^i_\mu = {\mathcal F}\!L^* \frac{\partial \phi_\mu}{\partial p_i}\,,
$$
where ${\mathcal F}\!L^*$ is the pull-back of the Legendre map ${\mathcal F}\!L$
from tangent space to phase space.
Therefore \bref{newas1} can be written as
\begin{equation}
\left(({\mathcal F}\!L^* \frac{\partial \phi_\mu}
{\partial p_i})\frac{\partial f^A}{\partial q^i}\right)_{\!\rm R}=
\left({\mathcal F}\!L^*\{\phi_\mu\,,f^A \}\right)_{\rm R} = 0\,,\label{assumpt2}
\end{equation}
which is equivalent to 
\begin{equation}
\hspace{6cm}
\{\phi_\mu\,,f^A \}_{\left\vert\!\!\begin{small}
\begin{tabular*}{\textwidth}{*{11}{c@{\hspace{1.7mm}}}c}
$ \phi_\mu=0 $
\\$ f^A=0$ 
\end{tabular*}\end{small}\right.} = \, 0 \,,\label{assumpt3}
\end{equation}
This is exactly \bref{assumpt}.

\subsection{Proof of  theorem 2}

Consider that the constraints $f^A =0$, holonomic and non-gauge fixing, generate, 
via Poisson brackets, a symmetry of the equations of motion. This means 
that the operator $\{-, f^A \}$ preserves the dynamics defined by the 
Dirac Hamiltonian $H_c + \lambda^\mu\phi_\mu$ ($H_c$ is the canonical Hamiltonian and
$\lambda^\mu$ are arbitrary functions)
\begin{equation}
\{H_c, f^A \} + \lambda^\mu \{\phi_\mu, f^A \} \approx 0\,,\label{can-stab}
\end{equation}
where the weak equality includes all the natural 
--that is, implied tby the theory itself--constraints of the theory. 

But the lhs in \bref{can-stab} is just the application of the dynamics to the 
constraints $f^A=0$ (in fact to the kinematical consequences of $f^A=0$, namely, 
$\dot f^A=0$). Since we already know that \bref{assumpt3} holds, the 
constraints $\chi^A$ determined before can also be given an alternative expresion 
originated in phase space
$$
 \chi^A = \left({\mathcal F}\!L^*\{H_c, f^A \}\right)_{\!\rm R}\,.
$$ 

Considering \bref{can-stab}, we infer that
$\chi^A$ is a combination of the natural constraints of the theory, specialized to 
the surface $f^A=0$. This means that in the reduced theory it is not
necessary to introduce the constraints $\chi^A$, 
because they will automatically be included 
among the natural constraints exhibited by the theory. The reduced theory is then 
formulated without additional constraints attached and therefore the truncation is 
consistent.

\bibliography{none}

\begingroup\raggedright\begin{thebibliography}{10}

\bibitem{Manton:1979kb}
{F}or an earlier reference see \\
N.~S.~Manton,
``A New Six-Dimensional Approach To The Weinberg-Salam Model,''
Nucl.\ Phys.\ B {\bf 158} (1979) 141.

\bibitem{Scherk:1979zr}
J.~Scherk and J.~H.~Schwarz, ``How To Get Masses From Extra
Dimensions,'' Nucl.\ Phys.\ B {\bf 153} (1979) 61.

\bibitem{Maccallum:gd}
M.~A.~MacCallum, ``Anisotropic And Inhomogeneous Relativistic
Cosmologies,'' In {\it   Hawking, S.W., Israel, W.: General
Relativity, an Einstein centenary survey. Cambridge University
Press, 1979, pags 533-580}.

\bibitem{pere-pep}
J.~M.~Pons and P.~Talavera,
``Consistent and inconsistent truncations: Some results and the issue of the
correct uplifting of solutions,''
Nucl.\ Phys.\ B {\bf 678} (2004) 427
[arXiv:hep-th/0309079].

\bibitem{Duff:hr}
M.~J.~Duff, B.~E.~Nilsson and C.~N.~Pope, ``Kaluza-Klein
Supergravity,'' Phys.\ Rept.\  {\bf 130} (1986) 1.

\bibitem{Cvetic:2003jy}
M.~Cvetic, G.~W.~Gibbons, H.~Lu and C.~N.~Pope,
``Consistent group and coset reductions of the bosonic string,''
Class.\ Quant.\ Grav.\  {\bf 20} (2003) 5161
[arXiv:hep-th/0306043].

\bibitem{Nastase:1999kf}
H.~Nastase, D.~Vaman and P.~van Nieuwenhuizen,
``Consistency of the AdS(7) x S(4) reduction and the origin of  self-duality
in odd dimensions,''
Nucl.\ Phys.\ B {\bf 581} (2000) 179
[arXiv:hep-th/9911238].

\bibitem{Witten:me}
E.~Witten,
``Search For A Realistic Kaluza-Klein Theory,''
Nucl.\ Phys.\ B {\bf 186} (1981) 412.

\bibitem{Duff:1983vj}
M.~J.~Duff, B.~E.~W.~Nilsson and C.~N.~Pope,
``Compactification Of D = 11 Supergravity On K(3) X T(3),''
Phys.\ Lett.\ B {\bf 129} (1983) 39.

\bibitem{plugging}
J.~M.~Pons,
``Plugging The Gauge Fixing Into The Lagrangian,''
Int.\ J.\ Mod.\ Phys.\ A {\bf 11} (1996) 975
[arXiv:hep-th/9510044].

\bibitem{dirac2} P. A. M. Dirac,
 ``Lectures on Quantum Mechanics,''
    (Yeshiva Univ.\ Press, New York, 1964)

\bibitem{bergm3}
J.~L.~Anderson and P.~G.~Bergmann,
``Constraints In Covariant Field Theories,''
Phys.\ Rev.\  {\bf 83} (1951) 1018.

\bibitem{strominger}
G.~T.~Horowitz and A.~Strominger,
``Black Strings And P-Branes,''
Nucl.\ Phys.\ B {\bf 360} (1991) 197.

\bibitem{chamseddine}
A.~H.~Chamseddine and M.~S.~Volkov,
``Non-Abelian solitons in N = 4 gauged supergravity and leading order  string theory,''
Phys.\ Rev.\ D {\bf 57} (1998) 6242 [arXiv:hep-th/9711181].

\bibitem{johmson}
C.~V.~Johnson,
``D-Branes,'' (Cambridge University Press, 2003)


\bibitem{chofreund}
Y.~M.~Cho and P.~G.~Freund,
``Nonabelian Gauge Fields In Nambu-Goldstone Fields,''
Phys.\ Rev.\ D {\bf 12} (1975) 1711.

\bibitem{mn}
M.~A.~Maldacena and C.~Nunez,
``Towards the large N limit of pure N = 1 super Yang Mills,''
Phys.\ Rev.\ Lett.\  {\bf 86} (2001) 588
[arXiv:hep-th/0008001].

\bibitem{arkady2}
G.~Papadopoulos and A.~A.~Tseytlin,
``Complex geometry of conifolds and 5-brane wrapped on 2-sphere,''
Class.\ Quant.\ Grav.\  {\bf 18} (2001) 1333
[arXiv:hep-th/0012034].

\bibitem{gubser}
S.~S.~Gubser, A.~A.~Tseytlin and M.~S.~Volkov,
``Non-Abelian 4-d black holes, wrapped 5-branes, and their dual  descriptions,''
JHEP {\bf 0109} (2001) 017
[arXiv:hep-th/0108205].

\bibitem{arkady}
L.~A.~Pando Zayas and A.~A.~Tseytlin,
``3-branes on spaces with R x S(2) x S(3) topology,''
Phys.\ Rev.\ D {\bf 63} (2001) 086006
[arXiv:hep-th/0101043].

\bibitem{paolo}
P.~Di Vecchia, A.~Lerda and P.~Merlatti,
``N = 1 and N = 2 super Yang-Mills theories from wrapped branes,''
Nucl.\ Phys.\ B {\bf 646} (2002) 43
[arXiv:hep-th/0205204].

\bibitem{cv}
A.~H.~Chamseddine and M.~S.~Volkov,
``Non-Abelian BPS monopoles in N = 4 gauged supergravity,''
Phys.\ Rev.\ Lett.\  {\bf 79} (1997) 3343
[arXiv:hep-th/9707176].

\bibitem{runing}
D.~Z.~Freedman, G.~W.~Gibbons and P.~C.~West,
``Ten Into Four Won't Go,''
Phys.\ Lett.\ B {\bf 124} (1983) 491.



\bibitem{Duff:1985cm}
M.~J.~Duff, B.~E.~W.~Nilsson and C.~N.~Pope,
``Kaluza-Klein Approach To The Heterotic String,''
Phys.\ Lett.\ B {\bf 163} (1985) 343.




\bibitem{Batlle:1985ss}
C.~Batlle, J.~Gomis, J.~M.~Pons and N.~Roman,
``Equivalence Between The Lagrangian And Hamiltonian Formalism For Constrained
Systems,''
J.\ Math.\ Phys.\  {\bf 27} (1986) 2953.


\end{thebibliography}\endgroup
\bibliographystyle{ssg}

\end{document}